# Mragyati[1]
## A System for Keyword-based Searching in Databases

**N.L. Sarda**  
Dept. of Computer Science and Engg.  
Indian Institute of Technology, Bombay  
Mumbai, India  

**nls@cse.iitb.ac.in**

**Ankur Jain**  
Dept. of Computer Science and Engg.  
Institute of Technology  
Banaras Hindu University,  
Varanasi, India  

**jain_ankur@excite.com**

## Abstract

The web, through many search engine sites, has popularized the keyword-based search paradigm, where a user can specify a string of keywords and expect to retrieve relevant documents, possibly ranked by their relevance to the query. Since a lot of information is stored in databases (and not as HTML documents), it is important to provide a similar search paradigm for databases, where users can query a database without knowing the database schema and database query languages such as SQL. In this paper, we propose such a database search system, which accepts a free-form query as a collection of keywords, translates it into queries on the database using the database metadata, and presents query results in a well-structured and browsable form. The system maps keywords onto the database schema and uses inter-relationships (i.e., data semantics) among the referred tables to generate meaningful query results. We also describe our prototype for database search, called Mragyati. The approach proposed here is scalable, as it does not build an in-memory graph of the entire database for searching for relationships among the objects selected by the user's query.

Index Terms : database search, keyword search, metadata, relevance ranking, web-based search

## 1. INTRODUCTION

The problem of extracting information scattered over the web is well known. Web search tools usually extract information pulled out from semi-structured documents. However, as information over web increasingly comes out of a database, it is crucial to provide searching of databases on the web directly. Since these databases serve applications that also update the data, it is not feasible to convert the database contents into a searchable (set of) HTML documents for use by search engines. A free-form search utility is required which can construct SQL queries from the keyword-based query (using metadata and other information from the database as transparently as possible) and present search results in a structured form. Furthermore, such a utility should be general so that it can be used with any database.

Figure 1 shows a popular way of implementing an application (often called as a gateway) for retrieval of data from a database over the web using technologies such as CGI and servlets. Data from the database can be exported to the web browser by using specially written programs. CGI (Common Gateway Interface), JavaScript, and SSI (Server Side Includes) programs are generally used for this purpose. These programs accept pre-defined user requests (usually through application-specific forms displayed in the browsers) and explicitly use knowledge about the database structure to extract the required data from the database. They execute application-specific (and appropriately parameterized) SQL queries for this purpose. Thus, these programs are specifically written for a given database. If the schema of the database changes, the search forms as well as the CGI database access programs also need to be changed. Another drawback is that a user cannot make free-form queries as (s)he generally does with a search engine. Often, these programs expose database schema in the result. The results are also unstructured, not permitting users to browse or drill-down into details as and if required.

---

[1] *Mragyati* is Sanskrit synonym for "search" or "hunt

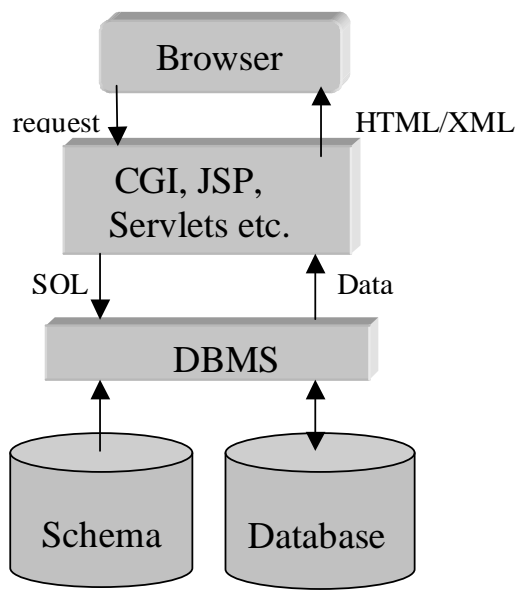

Figure 1 : Conventional implementation for data retrieval over the web

In this paper, we present our methodology for database searching on the web, and describe its prototype implementation called *Mragyati*. The paper is organized as follows. We take an overview of the related works on schema-independent and keyword-based database searching and search result presentation in Section 2. A brief overview of our approach in Section 3 is followed by a detailed presentation of our methodology in Section 4. This section describes query representation, classification, and execution. The relevance ranking of the retrieved results and the presentation of results in a hierarchical form are also discussed here. In Section 5, we give the architecture of our system, where we also discuss the internal databases required by our system. A prototype implementation named *Mragyati* is also described here. Section 6 concludes the paper with suggestions towards further extensions of this work.

## 2. RELATED WORK

Many organizations extract information from their databases and convert them into HTML pages. These pages are then published over the web. This approach is suitable for static data, as it requires the whole process to be repeated whenever the database contents change. The other approach, shown if Figure 1 and mentioned above, uses application-specific programs to run parameterized SQL queries and dynamically generate HTML pages containing required information. We have already mentioned limitations of this approach above.

There are some recent research efforts that address the problem of keyword searching in databases. We describe these efforts briefly below and then highlight main characteristics of our proposal.

Masermann and Vossen [5, 6] describe their research where the goal is to provide a simple, schema-independent web interface to relational databases, where queries, consisting of a few keywords (as in search engines), can be formulated in a declarative fashion. Whereas the standard SQL requires knowledge of tables and their attributes (i.e., schema definitions), they make use of parameters in SQL templates as placeholders for relations and attributes. When applied to a specific database, given its schema, they dynamically generate correct SQL statements from the parameters. These queries can then be executed, and results formatted appropriately for display to the user. The generation of queries is based on an SQL extension called Reflective SQL for handling procedural data. Thus, in their approach, user query is first translated into Reflective SQL, and the resulting expressions are then executed on the underlying database. They generate SQL queries based on a template that matches attributes of relations for the given keyword. The template is applied to a data dictionary that gives table names, column names and column types. Their approach is quite general and declarative. It only uses schema information, and does not build additional indexes or copy the database data (into a graph, etc.). No re-building of any intermediate data structures is required in case the database is updated by the application. However, there are certain limitations to their approach. They do not take into account other data semantics for generating more relevant results. They do not consider ordering results based on some notion of relevance. They also do not use vocabulary or any other information to relate keywords to the table attributes, but instead generate all possible queries based on only attribute types. Finally, the query results are in a pre-defined format that is difficult to understand and is not navigable.

Figure 2 shows the sample database (similar to one used in [5, 6]). For a simple search query containing "John", it produces the result shown in Figure 3. It can be seen that the result exposes schema of the database (by giving table/column names in the output), and its 'one format for all results' is too unstructured. It can be further seen that the result contains values like "***BO-3492***". If the user wishes to know more about it, he will have to fire a new query with "***BO-3492***". (Compare this with the result produced by Mragyati shown in Figure 4.) It is important to organize search results in an appealing and comprehensible form. This itself is an active area of research (see, for example [9]).

Shafer and Agrawal [7] propose a web-based interface and an application (called Eureka) for interactive exploration of databases, where, instead of forming specific SQL queries, the user browses through the data, places filtering predicates on attributes, or selects 'example' records for retrieving similar other records. Their interface seamlessly integrates continuous querying with result-browsing. The interface is intuitive and effective in applications (such as e-commerce) where users browse with a purpose. However, it lacks the power and appeal of keyword-based search that may transcend multiple tables. For good performance, they need to maintain complex data structures for caching results on the client side.

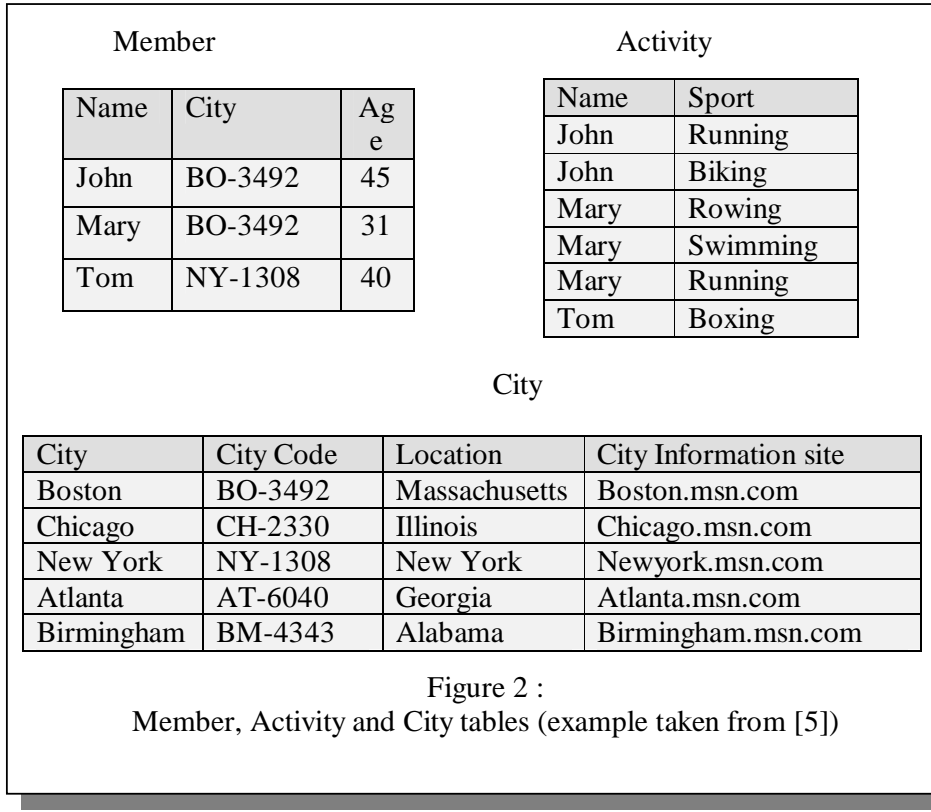

Figure 2 :
Member, Activity and City tables (example taken from [5])

| Table | Attribute | Value |
|---|---|---|
| Member | Name | John |
| Member | City | BO-3492 |
| Member | Age | 15 |
| Activity | Name | John |
| Activity | Sport | Running |
| Activity | Name | John |
| Activity | Sport | Biking |

Figure 3 : Result format in [5]

Goldman et al [4] extend the textual proximity search paradigm for searches within databases. A database is viewed as a graph with objects as nodes and relationships as edges. Relationships may be defined based on the structure or meaning of the database. They define proximity based on the shortest distance between the objects. The prototype allows queries to *find* objects of interest which are *near* to another set of objects. The objects found are ranked on their proximity.

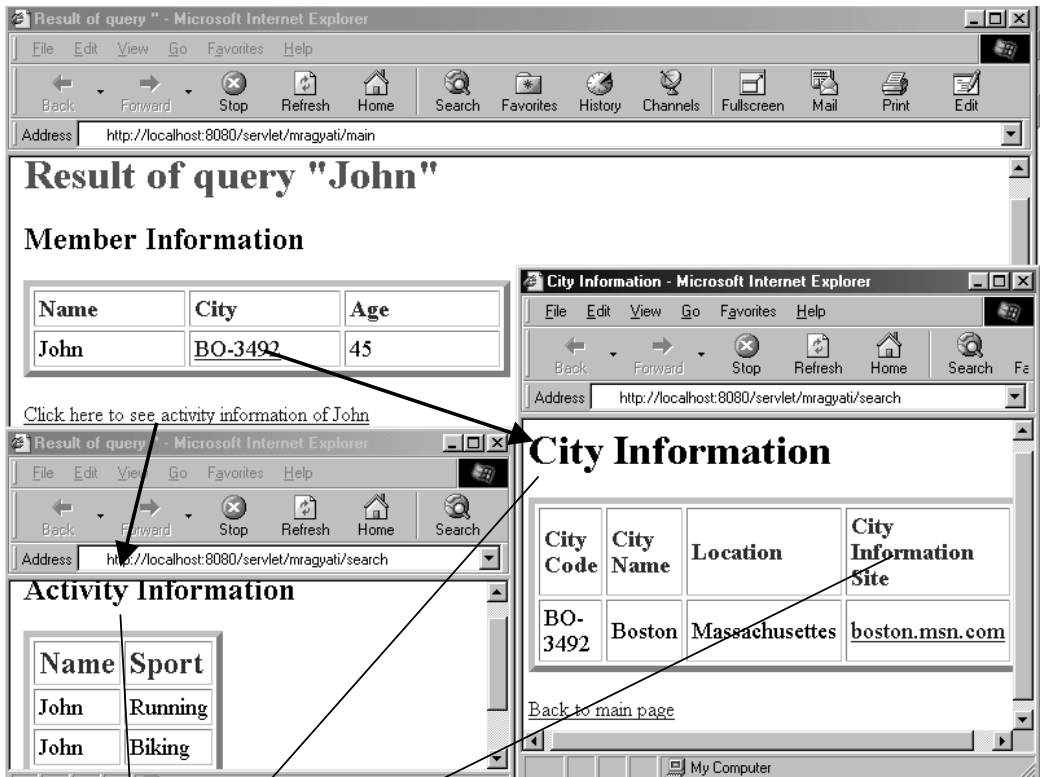

Figure 4 :
Screenshots produced by *Mragyati* on query "John"

Table and column names
taken from meta–database
description tables

Bhalotia [1] describe an extension to the work by Goldman et al [4] and give an efficient implementation for searches in relational databases. In their graph representation of a database, tuples are nodes, and edges capture foreign key and primary key relationships. They propose proximity based not only on shortest path, but also on the presence of other paths and in/out degrees of the nodes.

PESTO [2] by Carey et al is a tool for querying/browsing in object databases. It displays objects as windows on the screen, and facilitates moving through collection of objects and navigating along (reference) links between them. It supports 'query-in-place', by which filters can be specified for selecting objects as per a specific condition. It provides many useful features like 'synchronous browsing', last-query refinement, etc. The tool is schema-aware, and its search metaphor is not keyword-based searching.

Gey et al [3] identify a typical problem encountered by end-users in searching a database. The users are often not aware of how data are classified, categorized, abbreviated, named and represented in the database. They propose use of "vocabulary modules" to bridge the gap between the user's ordinary language, and the database system's metadata and stored data. They propose an agent-based architecture to develop domain specific vocabularies.

Toyama and Nagafuji [8] describes an extension to SQL for generating results as an HTML (or, another type of) document. The extension permits defining intra-page and inter-page hierarchical structures for browsing and navigation. The user must use SQL and the extensions to retrieve the required database objects. A keyword based search for the database contents is not the objective here.

In this paper, we describe our approach to keyword-based searching of databases. The distinguishing features of our approach are as follows :

i) it makes use of metadata internally to answer queries intelligently, and to retrieve related data in a navigable form. It provider a browser-based interface and allows queries in a form made popular by the search engines.
ii) We do not need to represent the entire contents of the database into a searchable structure (such as a graph used in Goldman et al [4] and Bhalotia [5] ). This is a very significant advantage for two important reasons. Our approach can scale to very large databases (where main-memory graph representation may not be possible), and databases may undergo concurrent updates without requiring update to or rebuilding of intermediate representations. Instead of searching for paths in the graph for the database, we generate and execute SQL queries, for which the relational database engines are very efficient.
iii) We make use of metadata (i.e., schema information) so that user queries may contain not only database values (such as "CS444", the course identifier for, say, the DBMS course) but also data classes and categories (such as "CS444 students").
iv) Our approach permits presentation of results ranked by their relevance to the query based on data inter-relationships. We propose relevance based on number of tuples retrieved by a query on related data, instead of tracing paths between related tuples. Thus, we use SQL queries internally to obtain results as well as their relevance. Ordering results based on their proximity to the main query result is available as an option. Usually, our system displays all related data in a structured and navigable form.
v) We present query results in more structured and navigable format compared to the approach by Masermann et.al [5, 6], where results are triplets of the form: (table name, column name, value). We also make use of vocabularies to map user terms (such as 'HINDI' with coded values stored in databases (such as 'HND'), and to provide meaningful substitutes for relation and column names in the metadata.
vi) Our approach is flexible for extending it to aggregate queries (e.g., "students ALL 'A'" – to retrieve students having all 'A' grades), and to queries that are best answered by taking joins of database relations. The approaches reported in literature query only one relation at a time, and may generate more queries (or, searches) than necessary.

## 3. OVERVIEW OF THE SYSTEM

In this section, we bring out the issues that need to be addressed in providing a keyword based search facility for databases. This will help us in identifying strategies and components required for the search system. We will use the database example shown in Figure 2, consisting of three relations, Member, Activity and City. To illustrate more complex examples, we will take the commonly familiar University example consisting of data about students, courses, teachers, course-registrations, publications by faculty, etc. We will also confine ourselves to the relational model for databases.

A DBMS manages data stored in multiple tables for an application. It also stores description of data in its internal catalogues. This description, also referred to as schema or metadata, consists of table and column specifications, and constraints such as primary and foreign keys, which capture inter-relationships among data. Application programs use SQL to retrieve and update data stored in the database. In a properly normalized database, the rows in tables describe business entities and transactions.

Our objective is to support a simple keyword based search in a database, where the user is not aware of the database schema. The queries here contain a set of keywords connected by Boolean operators (often, the AND operator is implied as in most search engines). The relative sequence of keywords is not important. In the context of database searching, a keyword may correspond to a value in some column in some table (in fact, it may be present in multiple columns in one/more tables). Examples of such queries are :
- "John"   : *meaning*, obtain all information about John
- "running Illinois" : *meaning*, get information (about members) having activity "running" and living in Illinois state

A keyword may also refer to data category or classification, which corresponds to table or column names in a database, as is the case in the following examples :
- "John activity" : *meaning*, get information about activities of John
- "Maths DBMS grades" : *meaning*, get grades of students from Maths department in the DBMS course (note: there could be multiple meanings when values can appear in multiple columns across tables)

The first step in query processing, quite obviously, is to interpret the keywords, by which the search system knows what the keyword refers to in the database.

To facilitate this interpretation of the keywords efficiently, the search system should have ready access to the metadata as well as the data. Since the metadata is usually stable and unchanging, it can be retrieved once and stored internally for efficient access (this may not be necessary if DBMS packages implement a standard metadata model such as OIM [10], or efficiently support access to the Information Schema defined in the SQL standard). To check whether a keyword occurs as a value of some attribute, one could do a brute force search of the entire database (or look under only the type-compatible columns in various tables), or build an inverse index to map values onto attributes. This value-attribute index may be selective, where only those columns whose values are likely to be used as keywords are included in the index. Typically, primary keys of tables representing business entities (such as member names, city names, etc.) are good candidates for indexing. The metadata extraction and building of value-attribute index needs to be done once, say, at the time of registering the application database with the search system.

Answering keyword queries for databases is more challenging that for text documents, where the given query is checked against one text document at a time. In a database, the keywords in a query may refer to one/more attributes in one/more tables, and these tables may or may not be related by a direct primary/foreign-key relationship. The search system may need to discover the relationships between the tables referred in the query so as to meaningfully answer the query (e.g., the "running Illinois" query above must also involve the member relation). Thus, the metadata should include key specifications for establishing these relationships.

Once the keywords are properly interpreted, the tuples that satisfy the keywords can be extracted from the individual tables. It is necessary to generate SQL statements for execution by the RDBMS for this purpose.

The query result may contain multiple tuples from one/more tables. These need to be organized in some order that may be meaningful for the user. The relevance or rank for each retrieved tuple may be defined using application-specific ranking function, or computed by the search system based on the relationships of the tuples with other tuples in the database.

Once the results are ordered, they can be displayed in a hierarchical format, permitting the user to *drill-down* into details if required. We can make use of the hyper-linking capabilities in the display in a browser-based interface.

The search system should take one more practical issue into consideration. The database objects (i.e., tables and columns) often have cryptic names, which are not known or easily understood by casual users. Also, values in the database are sometimes coded (such as 'M' for 'male'). It is essential to provide a proper translation from user's terminology to the internal names and values. The search system can build a 'vocabulary' to facilitate this translation.

In the next section, we elaborate on each of the above issues, and give details on how they are proposed to be handled in our search system.

## 4. SEARCH STRATEGY

### 4.1 Query Representation

A user query is a sequence of keywords connected by Boolean operators AND, OR and NOT. Often, the AND operator is implied. A keyword is generally a value expected to be found as value of some attribute in some table (it may be enclosed within quotes to demarcate it from other keywords in the query; we also assume that keywords do not contain wild characters). A keyword may also refer to metadata element, either an attribute name or relation name. We illustrate below a few example queries for the database shown in Figure 2.

        Q1 : John
        Q2 : John and Mary
        Q3 : John Activity
        Q4 : John Sport

In this paper, we consider queries that are conjunction of keywords. The scope of NOT is assumed to be confined to a single keyword (as in "not running") of value type (obviously, "not activity", where not is applied to a relation name, has little meaning). The approach presented here can be extended to remove the above restrictions.

A given query is initially parsed and represented as two connected trees before analyzing them for generating queries on the underlying database. The trees capture both the metadata and query terms. The top tree has maximum 4 levels as follows:

    Level  0 : root, represents the whole query
            1 : nodes here represent relations involved in the query

2 : nodes here represent attributes referred to in the query
3 : nodes here give keywords that are values of the attributes.

The parsing and processing of keywords in the query facilitates construction of the above tree structure. The processing here accesses schema information and also the values-to-attribute index. The trees for query Q1 and Q4 above are shown below:

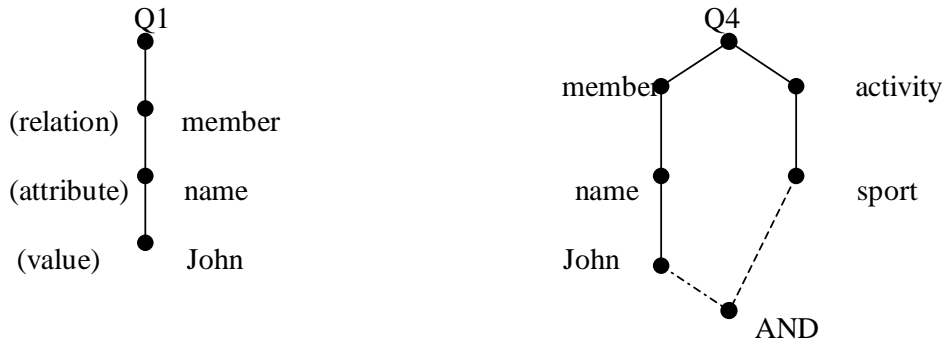

Note that some of the lower levels in the tree may be absent (when the query specifies attribute/table names).

The second tree represents the Boolean expression of terms by linking the leaf nodes in the first tree by Boolean operator nodes (e.g., in the tree for query Q4 above, we define an 'AND' node between its two keywords). Since we are considering only conjunctive queries, the second tree will not be mentioned further.

## 4.2 Query Classification

The query tree basically provides an interpretation for each keyword. It contains a path from the root to each keyword. A full path consists of relation name (R), attribute name (A) and value v (optionally preceded by 'not'). In general, however, attribute name as well as the value may be absent. Thus, a path may take one of the following forms (referred as 'path term') :
- (R, A, v)
- (R, A)
- (R)

A query, then, is a conjunction of path specifications of the above forms.

From the processing point of view, we classify the queries into the following categories:

i) single-relation query : it contains only one relation in its path terms (e.g., (member, name, 'John')). If it contains multiple path terms, they are on different attributes of the same relation.
ii) multiple-relation query: it contains two or more path terms, each involving a separate relation. An example query would be ((activity, sport, biking), (city, location, Georgia)).
iii) repeated-attribute query: the path terms in this class contain same (relation, attribute) pairs two or more times. An example query will be ((activity, sport, running), (activity, sport, biking)), where the user would be interested to know objects (members, in this case) that relate to both these sports.

## 4.3 Query Processing

We now describe the processing for each type of query. Note that the search system uses database metadata, and database contents (not directly, but through a value-attribute mapping index, as described in the next section) to parse, translate and classify the query. In the query processing phase, where it obtains query results, the system will further use metadata and access database contents. The metadata is used to identify relationships among data. Various queries (in SQL) generated in this phase are submitted to the backend DBMS for obtaining results. The processing performed for each query category is described below.

i) Single relation query: This is the simplest query to translate in SQL and execute. (R, A, v) query specification is translated into

       **select** *
       **from** R
       **where** R.A = v

If values are specified for additional attributes of R (all being distinct), more predicates are included in the where-clause. The query result along with schema specification for R is handed over first to the relevance ordering module (if requested) and then to the result display module. (Note that the where-clause is not required when the path term does not contain attribute and/or value specification).

ii) Multiple-relation query: such a query contains two or more path terms, each on a different relation. The terms are first ordered based on key relationships between their target relations. Consider two related terms ($R_1$, $A_1$, $v_1$), ($R_2$, $A_2$, $v_2$) in the query with a foreign key relationship between them such that the primary key K of R1 is a foreign key F in R2 (i.e., there is 1-to-many relationship between R1 and R2). The query processor, in this case, must ensure that the tuples retrieved from R2 are only those that have key values same as keys in the selected tuples of R1. The SQL query generated for retrieval from R2 will include this key condition also:

       **select** *
       **from** R2
       **where** R2.A2 = v2 **and**
          R2.F **in** (**select** R1.K **from** R1
                    **where** R1.A1 =v1)

Similarly, the SQL query generated for R1 will also contain a condition on key K so that we do not retrieve those tuples of R1 which do not have counterparts in tuples selected from R2 :

       **select** *
       **from** R1
       **where** R1.A1 = v1 **and** R1.K **in**
          (**select** R2.F **from** R2 **where** R2.A2 = v2)

As an example, consider the query ((dept, location, London), (emp, job-type, Programmer)), where we expect to obtain Programmer employees working in departments located in London. It can be readily seen that key predicates should be added to the SQL statements as described above.

When the query contains path terms for many relations, the above approach can be generalized to include key constraints in the generated SQL statements for all the relations related by key relationships.

Now consider a query ((R1, A1, v1), (R2, A2, v2)), where R1 and R2 are not directly related by a key relationship. The query processing here examines the metadata to find intermediate relation(s) to relate them. Without loss of generality, let R3 (which may actually be a join of 2 or more relations) be an intermediate relation that relates to both R1 and R2 through primary/foreign key specifications. We generate SQL statements on R1 and R2 by using key relationships with R3. Let us consider the query: ((Course, dept, CS), (Student, status, weak)), where the relations course (with key CNO) and student (with key SNO) are related through the Study relation having both CNO and RNO as foreign keys. We can now generate SQL statements that will give us CS-*courses* having *weak students*, and *weak students* from CS *courses*. The query on course will be:

    **select** * **from** Course C
  **where** dept = 'CS' **and**
      exists (**select** * **from** Student S, Study T
        **where** S.sno = T.sno **and**
        T.cno = C.cno **and**
        S.status = 'weak')

The processing here passes multiple result sets to the relevance ordering module (if required) and then to the result display module.

iii) Repeated-attribute query: These queries require a separate treatment for two reasons: firstly, we can not perform 'AND' on these predicates (as that would give a null result; this can't be the intension of the user); secondly, we must examine other relations that have 'many' relationships with this relation so that we can retrieve objects related to all the objects specified by the keywords. We will illustrate the processing here by the following example query:

  (Faculty, name, Sudarshan), (Faculty, name, Soumen),
    (Papers, area, QueryOptimization)

This query will retrieve the two faculty and their (joint) research papers in the area of query optimization. The name attribute in the relation Faculty is repeated. The query processing here analyzes the metadata to deduce many-to-many relationship between faculty and Papers (the key of papers is a composite key containing faculty-id fid). The SQL query generated on Papers will contain predicates to ensure relationship with both faculty :

  **select** * **from** Papers p
  **where** p.area = "QueryOptimization" **and**
    exists (**select** *
      **from** Papers $q_1$, Papers $q_2$

```
        where q₁.pid = p.pid and
              q₂.pid = p.pid and
              q₁.fid = (select fid from Faculty
                        where name = "Sudarshan)
              and
              q₂.fid = (select fid from Faculty
                        where name = "Soumen"))
```

Note than the pattern in the inner query can be easily repeated in case the attribute (name, in this case) was repeated in user query more than twice.

The SQL query generated for the relation having repeated attribute will contain OR operator in its predicate, as shown below for the faculty table:

```
select * from faculty
where  name = "Sudarshan"  OR name = "Soumen"
```

## 4.4 Relevance Ordering

An optional feature in Mragyati system is to get the objects (i.e., tuples) in a result set ordered by some implicit or explicit ranking function. We will mention here the implicit ranking based on foreign-key relationships, and a simple explicit ranking based on application-defined sort order.

Let R be a set of tuples selected for display. The metadata for R indicates whether its primary key K is a foreign key F in another relation S. The tuples in R are then ordered by the count of tuples in S to which they are related. This is achieved by generating an SQL to count related tuples and to order R :

```
          select R.*, C as (select count(*)
                             from S
                             where S.F = R.K)
          from  R
          order by  desc  C
```

If R is related to more than one relation, the count C above can be the sum of counts of related tuples in the various relations.

Optionally, an application-specific ordering can be used. The application may record preferred ordering as data in a pre-designated table. The table can indicate attributes and sort orders to be used in ranking. For example, the salary attribute of Emp relation (in descending order) may be used to order employees in the result display.

## 4.5 Result Display

The result of a query consists of one or more tuple sets. A tuple set is a set of tuples from a single relation R. The result includes one set for each relation present in the (translated) query. The query

((course, dept, CS), (student, status, weak))

will contain two tuple sets, one from course table and the other from student table. Each tuple set, along with the metadata of its stored relation is passed to the result display module. Each tuple set is displayed in a separate frame.

Given a tuple set from R for display, the display is defined as follows:
 i) the selected tuples with their values are displayed (by their rank, if required)
 ii)  if attribute A in R is a foreign key, values of attribute A are highlighted, and links are generated to fetch the target tuples (from another table where it is primary key), and
 iii)  if attribute B in R is its key, and B is a foreign key in another table S, and S is not already included in the result, a line is generated below the R tuple to provide a hyperlink to the related tuples in S.

Figure 4 shows the result for query "John", which produces a single tuple set from Member relation, and the set contains only one tuple. The city attribute is highlighted as a URL to link to City information, and there is a line (made self-explanatory by using vocabulary of the application) added below the displayed tuple to get all activity information.

## 5. SYSTEM ARCHITECTURE

### 5.1 Main Modules

We now describe the software architecture of our search system for supporting keyword searches on databases. We will follow this with description of our current implementation (which we have named as Mragyati).

The main building blocks of the search system are shown in Figure 5. An existing database application is first registered with the system. The initialization (registration) module builds the metadata by consulting the database catalogue. The module can interact with the application administrator to build a vocabulary whereby meaningful descriptions are associated with the often-cryptic table and column names found in database schema. The vocabulary also includes expansion for codes occurring as values in the database (e.g., 'F' for 'female', 'CS' for 'Computer Science', etc.). Often, a database contains table for codes and their expansions. These tables can be made a part of the system vocabulary. Finally, the registration module builds a 'value-attribute' index, whereby the search system can map a given value to one or more attributes in the database. The entries in this index have the form <value, attribute, relation> (an example will be (John, name, Member)). This index is useful for quickly translating a keyword in user's query into its metadata. It may be noted that the application administrator may choose columns from the database tables for this index. It is not mandatory that every individual value occurring in the database be indexed.

The system provides a browser-based interface. The user supplies a query string and invokes the search system. The query results are also displayed in the browser as an HTML document.

The user query is analyzed and translated by the Query Parsing module. The module uses application vocabulary to translate user's terms into internal values wherever necessary (thus, the query 'Computer Science female students' will be converted into 'CS F Students').

Next, it maps terms in user's query onto database metadata by using the value-attribute mapper module. It uses the information provided by the mapper module to construct the query tree. The parsing module may also call the metadata manager module to check if a term in the query is a table/column name (instead of a value).

The translated query is sent for evaluation. The evaluation module makes use of metadata to convert the query path terms into SQL statements as described in the previous section. The metadata module primarily provides primary/foreign key information among tables, and about tables that may have key relationships with a given table. The generated SQL queries are sent to the DBMS (managing the application database) for execution, and the returned results are collected as tuple sets. The results are next sent to the relevance-ordering module, which defines ordering of tuples in each set. Finally, the (ordered) tuple sets with their description are sent to the display module for generating HTML documents with appropriate hyper links.

## 5.2 Search System Database (SS-DB)

The search system builds up its own information base when an application database is initially registered with it. This information base contains metadata, value-attribute mappings, and application vocabulary as described above. In this section, we briefly describe how this information is structured for storage in a database system so that it can be efficiently queried.

The value-attribute mapper builds a table called VMAP consisting of columns for value, attribute name, and relation name. If a value in application database appears in more than one table, VMAP will contain a separate entry for each. VMAP is indexed on value for efficient access.

The application vocabulary is also represented by a table called VOC with two columns : one for internal value and one for its external description. It is indexed on external description column.

The application metadata contains description of tables in the application database. The metadata itself is stored in the search system as a database consisting of the following four tables:

i) Table description (TBL) : This table contain names and descriptions of tables in the application database. Each table is also given an internal identifier T-id. The description is a text providing meaningful explanation for the table contents (as provided by the application administrator during the registration).

ii) Column description (COL) : This table contains the following columns : T-id, column-id (internally generated), column name, type, and description (as provided by the application administrator).

iii) Primary key table (PKEY) : This table contains primary key and unique attributes (candidate keys) for application tables. Since a table may have a composite (i.e., multi-attribute) key, the keys are given internal numbers for each table. This table has the columns : (T-id, K-no, column-id). The primary key for a table is given K-no as 1. If a key contains two attributes, the key table contains two entries with same K-no.

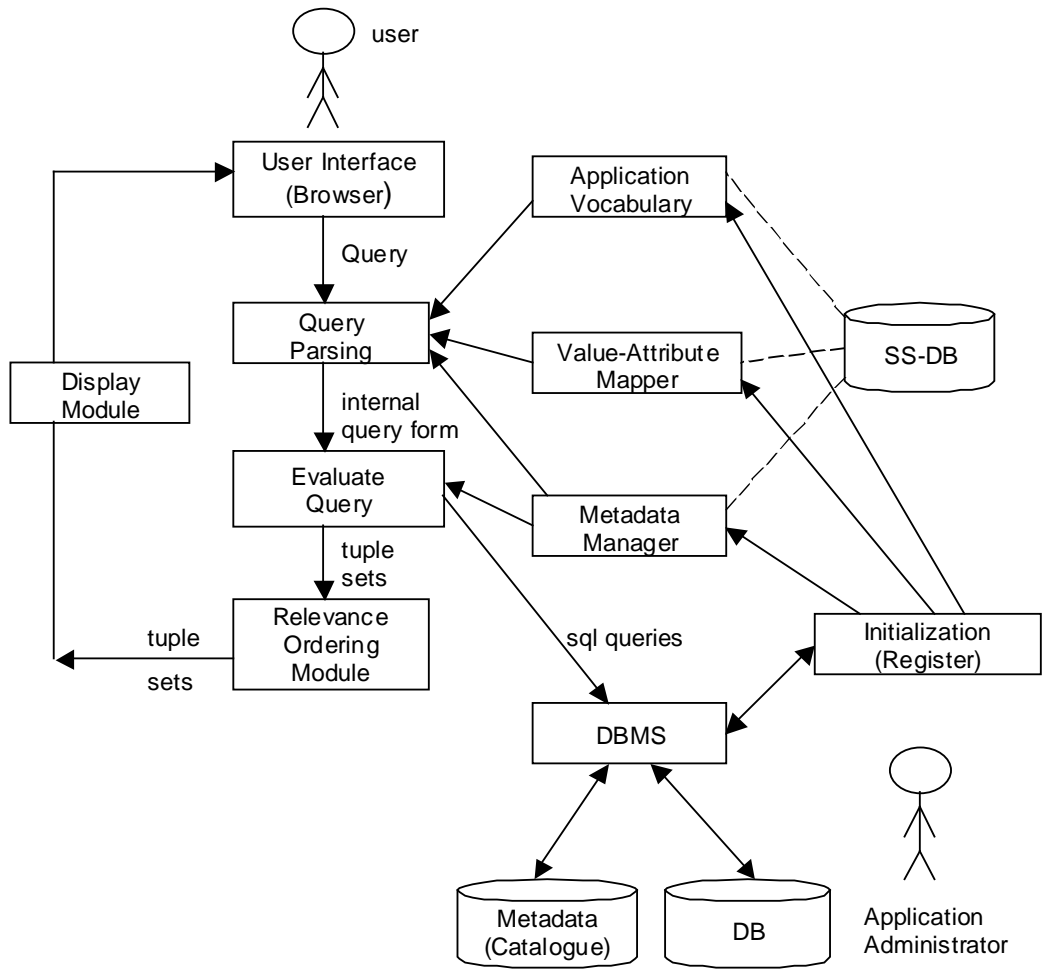

Figure 5 Architecture of the rearch system showing main modules and flow between them

iv) Foreign key table (FKEY) : This table contains foreign key specifications associated with database tables. It relates a foreign key (which could be a composite key) in one table with a candidate key in another table. It has the following contents: (T1-id, F-no, column-id1, T2-id, K-no, column-id2), where T1-id identifies table containing foreign key and T2-id identifies the table containing the candidate key. FKEY table contains multiple entries (with same F-no and K-no) for multi-attribute foreign and candidate keys.

The metadata module also analyzes inter-relationships between database tables and stores them for ready reference of the evaluate-query module. If R1 and R2 are relations in the application's database schema, they are related directly if they have a primary/foreign key relationship between them (say, given by F-no value $f_{12}$ in the table FKEY above). Relations R1 and R2 may be related by a chain of relations R3, R4, …, Rk. The application schema then has a path R1, R3, …, Rk, R2, where every successive pair is related by a primary/foreign key relationship. The metadata module pre-computes these paths and stores them in PATH table, where each path is simply stored as a string (R1, $f_{13}$, R3, $f_{34}$, R4, ….). Note that more than one path may exist between two relations.

## 5.3 Mragyati : the prototype implementation

Mragyati is a prototype implementation of the search system proposed in this paper. It is implemented using Java servlets, JSP, and JDBC for database access. It uses Oracle 8.1.5 to stores its own database (SS-DB) and it has been tested for Oracle-based application databases.

When an application is registered with Mragyati, it uses JDBC metadata calls to obtain metadata (such as table names, column names, etc.) of the application database. It also permits associating meaningful descriptions with table/column names as shown in Figure 6. It allows the application administrator to mark tables and columns for loading into the vocabulary (VOC) and value-attribute mapping (VMAP) tables. In the current implementation, we do not analyze schema to build the PATH table.

Mragyati accepts queries as a sequence of keywords, and maps them onto the application metadata as described earlier. It presently implements query processing for only the first two query categories (repeated-attribute queries are not implemented). Also, Mragyati currently does not discover indirect relationships between tables involved in the query (for which the PATH table would be needed). The SQL queries are generated from path terms, and submitted for execution using JDBC calls. The results are submitted to the display module (where JSPs for handling hyperlinks in the display results are also generated). Table names are also displayed as URL. Clicking on a table name displays the contents of the table. Since Java supports multi-threading, Mragyati accelerates the search process by executing SQL queries on multiple threads whenever possible. Currently, we do not implement the Relevance Ordering module. Figure 4 shows a typical result displayed by Mragyati.

Since all values present in the application database may not be selected for mapping in the VMAP table, a situation may arise when Mragyati cannot interpret a keyword present in user's query. This situation can be handled using one of the following two approaches:
i) return a message to the user about system's inability to handle the keyword, and
ii) undertake a search for the keyword under every (type-compatible) column of every table to dynamically check if the keyword occurs as a value anywhere in the application database.

Mragyati presently implements the first approach, although the second approach has some important advantages. At the cost of slow (and potentially exhaustive) search at the execution time, the second approach can incrementally maintain VMAP table and avoid periodical re-registration of the application database (due to application level updates).

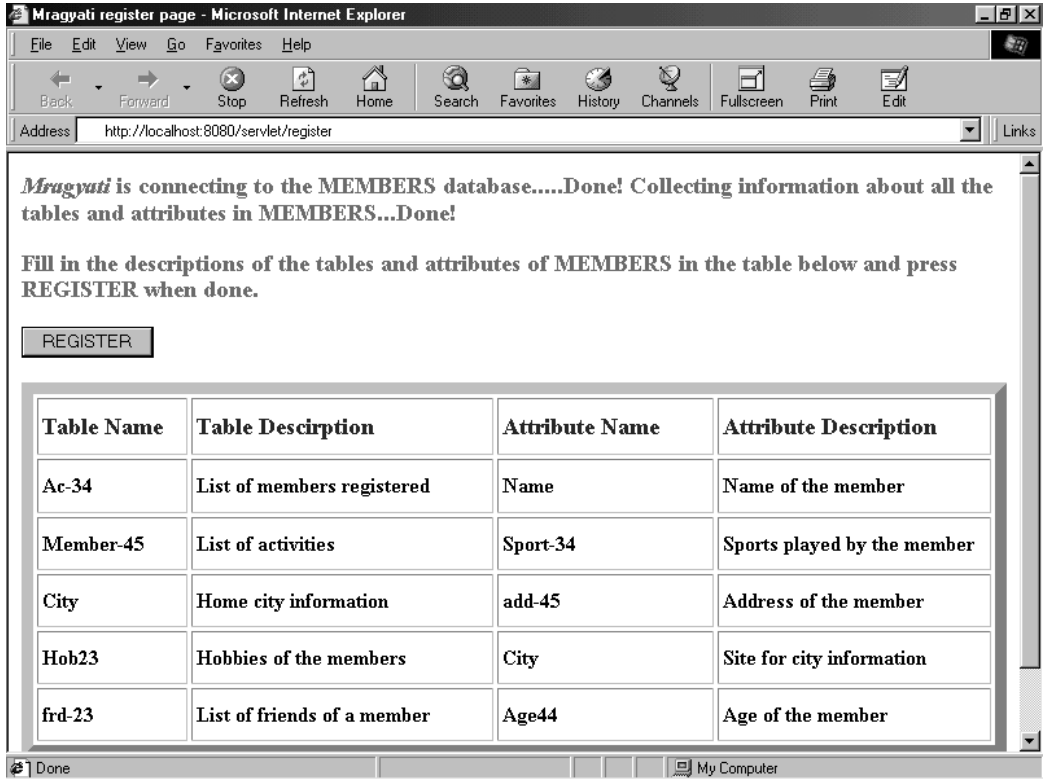

Figure 6 : Mragyati's interface for associating meaningful
Descriptions with table and column names

## 6. CONCLUSIONS

Information on the web increasingly comes from databases maintained by the organizations, which use technologies such as CGI, JSP/ASP, Servlets, etc. to pull out and transform the required data for viewing within the browsers. The interfaces in these cases are pre-defined, and the applications use the database metadata implicitly by hard-coding the metadata in the application. There is clearly a need to support ad hoc and keyword-based database search paradigm made popular by the search engines. The database schema is transparent here to the user.

In this paper, we have presented design and prototype implementation of such a tool. It makes use of the database structure and semantics to intelligently answer user queries consisting of a few keywords. Our approach offers many advantages compared to earlier approaches, particularly those in Masermann et al [5, 6], Goldman et al [4] and Bhalotia [1]. Firstly, we do not need to copy database contents into an intermediate and in-memory structure (i.e., graph) for searching, thus making the approach efficient and scalable. We generate SQL statements to retrieve the required data objects (which generated SQLs can run concurrently with the normal application operations), and fine-tune the statements based on inter-relationships between the data. The results generated by our tool are hierarchically structured and browsable.

The prototype, called Mragyati, supports most of the features proposed here. It can be easily extended to include relevance ranking of the retrieved tuples, and for additional type of queries. A useful feature could be to display results, wherever possible, by taking joins of the tables referred in the query. This could avoid fragmentation in the result, thereby reducing unnecessary browsing (in fact, we can present results in nested table format as in [8]).

We plan to explore two interesting research directions as part of the future work. The first relates to ranking and proximity analysis between the database objects in the result. This need not be based only on cardinalities of relationships among the objects (or, number of paths as in [4, 1]), but also on application specific ranking functions. For example, employees may be ranked by salary, years of service, or a combination of such factors. The other research area would be to allow more naturalness in queries by using NLP(Natural Language Processing) techniques. For example, a query such as "Sudarshan's students" is more natural and more precise in its intent that the query "Sudarshan students", where we need to analyze all possible relationships between Sudarshan (a faculty) and students, including the case of students named as Sudarshan!

## **References:**

1. G. Bhalotia, 'Keyword searching in Databases using BANKS', B.Tech. project report, I.I.T. Bombay, April 2001.
2. M. Carey, L. Haas, V. Maganty and J. Williams, 'Pesto : An Integrated Query/Browser for Object Databases', Proc. VLDB '96, Bombay, p. 203-214, 1996.
3. F. Gey et.al., 'Advanced Search Technologies for Unfamiliar Metadata', Third IEEE Meta-data Conference (Meta-Data 99), April 6-7, 1999, Bethesda, Maryland
4. R. Goldman, N. Shivakumar, S. Venkatasubramanian, H. Garcia-Molina, 'Proximity Search in Databases', Prof. VLDB '98, pp.26-37, 1998.
5. U. Masermann and G. Vossen, 'Design and Implementation of a Novel Approach to keyword Seraching in Relational Databases', Prof. Of ADBIS-DASFAA Symp. On Advances in Databases & Information Systems, Sept. 5-8, 2000, Prague, Czech Republic
6. U. Masermann and G. Vossen, 'SISQL : Schema Independent Database Querying (on and off the Web)', Proc. Of IDEAS2000, Sept. 18-20, 2000, Yokohoma, Japan
7. J.C. Shafer and R. Agrawal, 'Continuous Querying in Database Centric Web Applications', Proc. Of 9th Intl. WWW Conference, Elsevier Science, May 15-19, 2000, pp. 519-531
8. M. Toyama and T. Nagafuji, 'Dynamic and Structured Presentation of Database Contents on the Web', Proc. EDBT'98 (Spain, 1998), LNCS # 1377, Springer-Verlag, 1988, pp. 451-465.
9. J. Cugini, 'Presenting Search Results: Design, Visualization, and Evaluation', Information Technology Laboratory, Information Doors Workshop at the ACM HyperText 2000 Conference (also available at http://www.itl.nist.gov/iaui/vvrg/cugini/irlib/paper-may2000.html)
10. Meta Data Coalition, 'Open Information Model, version 1.0', August 1999, available from http://www.mdcinfo.com.